\documentclass[12pt,preprint]{aastex}
\usepackage{lscape}

\def\arcsec{\ifmmode '' \else $''$\fi}

\def\arcsecpoint{\ifmmode ''\!. \else $''\!.$\fi}

\def\kms{\ifmmode {\rm km\ s}^{-1} \else km s$^{-1}$\fi}
\def\Msun{\ifmmode {\rm M}_{\odot} \else M$_{\odot}$\fi}
\def\Lsun{\ifmmode {\rm L}_{\odot} \else L$_{\odot}$\fi}
\def\Zsun{\ifmmode {\rm Z}_{\odot} \else Z$_{\odot}$\fi}
\def\ergsAcm{erg\,s$^{-1}$\,cm$^{-2}$\,\AA$^{-1}$}
\def\ergscm2{erg\,s$^{-1}$\,cm$^{-2}$}
\def\icm3{{\rm cm}^{-3}}
\def\icm2{{\rm cm}^{-2}}
\def\qo{\ifmmode q_{\rm o} \else $q_{\rm o}$\fi}
\def\Ho{\ifmmode H_{\rm o} \else $H_{\rm o}$\fi}
\def\ho{\ifmmode h_{\rm o} \else $h_{\rm o}$\fi}
\def\ltsim{\raisebox{-.5ex}{$\;\stackrel{<}{\sim}\;$}}

\def\vFWHM{\ifmmode v_{\mbox{\tiny FWHM}} \else
            $v_{\mbox{\tiny FWHM}}$\fi}
\def\CCF{\ifmmode F_{\it CCF} \else $F_{\it CCF}$\fi}
\def\ACF{\ifmmode F_{\it ACF} \else $F_{\it ACF}$\fi}
\def\Halpha{\ifmmode {\rm H}\alpha \else H$\alpha$\fi}
\def\Hbeta{\ifmmode {\rm H}\beta \else H$\beta$\fi}
\def\Hgamma{\ifmmode {\rm H}\gamma \else H$\gamma$\fi}
\def\Hdelta{\ifmmode {\rm H}\delta \else H$\delta$\fi}
\def\Lya{\ifmmode {\rm Ly}\alpha \else Ly$\alpha$\fi}
\def\Lyb{\ifmmode {\rm Ly}\beta \else Ly$\beta$\fi}
\def\Lyg{\ifmmode {\rm Ly}\beta \else Ly$\gamma$\fi}

\def\hei{He\,{\sc i}}

\def\cii{C\,{\sc ii}}
\def\ciii{\ifmmode {\rm C}\,{\sc iii} \else C\,{\sc iii}\fi}
\def\civ{\ifmmode {\rm C}\,{\sc iv} \else C\,{\sc iv}\fi}

\def\oi{O\,{\sc i}}

\def\o5007{[O\,{\sc iii}]\,$\lambda5007$}

\def\mgii{Mg\,{\sc ii}}

\def\siiv{Si\,{\sc iv}}
\def\siIV{Si\,{\sc iv}}

\def\siII{Si\,{\sc ii}}

\def\feii{Fe\,{\sc ii}}
\def\feiii{Fe\,{\sc iii}}

\def\alii{Al\,{\sc ii}}
\def\aliii{Al\,{\sc iii}}

\def\niII{Ni\,{\sc ii}}

\begin{document}

\title{Quasar Outflow Contribution to AGN Feedback: Observations of QSO SDSS~J0838+2955}

\author{Maxwell Moe\altaffilmark{1}, Nahum Arav\altaffilmark{2}, Manuel A. Bautista\altaffilmark{2}, and Kirk T. Korista\altaffilmark{3}}

\altaffiltext{1}{Department of Astrophysical and Planetary Sciences and CASA, University of Colorado, 389-UCB, Boulder, CO 80309, USA}

\altaffiltext{2}{Department of Physics, Virginia Polytechnic Institute and State University, Blacksburg, VA 24061, USA}

\altaffiltext{3}{Department of Physics, Western Michigan University, Kalamazoo, MI 49008-5252, USA}

\begin{abstract}

We present a detailed analysis of the Astrophysical Research Consortium 3.5m telescope spectrum of QSO SDSS~J0838+2955.  The object shows three broad absorption line (BAL) systems at 22,000, 13,000, and 4900 \kms\ blueshifted from the systemic redshift of z=2.043.  Of particular interest is the lowest velocity system that displays absorption from low-ionization species such as \mgii, \alii, \siII, \siII*, \feii\ and \feii*.  Accurate column densities were measured for all transitions in this lowest velocity BAL using an inhomogeneous absorber model. The ratio of column densities of \siII* and \feii* with respect to their ground states gave an electron number density of log $n_e$ (cm$^{-3}$) = 3.75 $\pm$ 0.22 for the outflow.  Photoionization modeling with careful regards to chemical abundances and the incident spectral energy distribution predicts an ionization parameter of log $U_H$ = -1.93 $\pm$ 0.21 and a hydrogen column density of log $N_H$ (cm$^{-2}$) = 20.80 $\pm$ 0.28.  This places the outflow at 3.3$^{+1.5}_{-1.0}$ kpc from the central active galactic nucleus (AGN).  Assuming that the fraction of solid angle subtended by the outflow is 0.2, these values yield a kinetic luminosity of $(4.5^{+3.1}_{-1.8})\times10^{45}$ erg~s$^{-1}$, which is (1.4$^{+1.1}_{-0.6}$)\% the bolometric luminosity of the QSO itself. Such large kinetic luminosity suggests that QSO outflows are a major contributor to AGN feedback mechanisms.

\end{abstract}


\section{Introduction}
\label{sec:introduction}

Quasar outflows have been increasingly invoked as a primary feedback mechanism to explain the formation and evolution of supermassive black holes, their host galaxies, the surrounding intergalactic medium (IGM), and cluster cooling flows \citep{Elvis2006}.  Several of these feedback mechanisms require that the kinetic luminosity ($\dot{E}_k$) of the outflowing material would be $\sim$5\% the bolometric luminosity of the QSO itself assuming the active galactic nucleus (AGN) is accreting near the Eddington limit \citep{Scannapieco2004}. The kinetic luminosity of absorption outflows (observed as blueshifted troughs in the quasar spectrum) is directly proportional to the distance $R$ of the outflow from the central AGN, the total hydrogen column density $N_H$, and the fraction $\Omega$ of the solid angle subtended by the wind (see Section 5).  These parameters must therefore be accurately determined in order to infer the significance of such outflows for AGN feedback.  

Previous attempts to measure $\dot{E}$$_k$ \citep{Wampler1995, deKool2001, deKool2002, deKool2002b, Hamann2001} had large statistical uncertainties \citep[typically more than an order of magnitude, see ][]{Dunn2009} and suffered from unquantified systematic errors due to the use of inadequate absorption models. The latter lead to unreliable measurements of column densities in the observed troughs ($N_{ion}$), which are crucial for determining almost every physical aspect of the outflows. To determine $N_H$ we use photoionization modeling to find the ionization equilibrium that can reproduce the measured $N_{ion}$. To find $R$, in addition to the ionization equilibrium, we also need the number density of the gas which can be determined by comparing the ionic column densities of excited metastable states to their ground states \citep{Wampler1995,deKool2001,Korista2008}.  We assume $\Omega$ = 0.2, as is the case for broad absorption line (BAL) quasars in general \citep{Hewett2003}.  We caution that $\Omega$ is not directly constrained by our observations, and therefore express the kinetic luminosity in terms of $\Omega_{0.2}$ (see Section~5).

In order to determine accurate  $\dot{E}_k$ in quasar outflows, we launched a multistage research program. We first developed analysis techniques that allow us to determine reliable trough column densities \citep[][and references therein]{Arav2008}. We then performed an exhaustive search through $\sim$50,000 QSOs with redshift $z>$0.25 and Sloan Digital Sky Survey (SDSS) magnitude $r<$19.3 in the SDSS Data Release 6 \citep{Adelman2008}; $\sim$100 objects were found that exhibited absorption systems containing transitions from excited metastable states of \hei*, \siII*, \feii*, \feiii* and/or \niII*.  Of these objects, approximately one third exhibited only \hei* (E=159,856 cm$^{-1}$) without transitions from other energy levels nor other ions in metastable states that could serve as density diagnostics.  Of the remaining $\sim$70 QSOs, $\sim$30 objects displayed the intrinsic absorption system containing the excited states at velocities $\gtrsim$1,000 \kms\ with respect to the systemic restframe of the QSO and continuum flux levels $\gtrsim$10$^{-16}$ \ergsAcm\ in the observed frame.

Here we investigate one of the brightest objects from this list with wide enough absorption troughs to allow sufficiently accurate extraction of column densities with a moderate resolution spectrum.  QSO SDSS~J083817.00+295526.5 (hereafter, SDSS~J0838+2955) exhibits three systems of BALs with the lowest velocity system at -4,900 \kms\ containing absorption from \siII\ and \siII* $\lambda\lambda$1527,1533 in the SDSS spectrum.  With an SDSS r magnitude of 17.85 and a redshift z=2.043, this object was ideal for follow-up observations because the optical portion of the spectrum bracketed both high ionization species of \civ\ and \siIV\ as well as the low ionization species of \cii, \mgii, \alii, \aliii, \siII, and \feii.  The plan of this paper is as follows.  In Section~\ref{observations}, we detail the observations and data reduction of SDSS~J0838+2955.  Gas column and number densities are measured in Section~\ref{column} after determining the inhomogeneity of the absorber across the continuum source.  Photoionization modeling of the outflow with careful regards to the incident spectral energy distribution (SED) is presented in Section~\ref{photoion}.  We discuss the results for distance and kinetic luminosity in Section~\ref{results}.  For this analysis, we adopt a standard $\Lambda$CDM cosmology with H$_0$=70 \kms\ Mpc$^{-1}$, $\Omega_m$=0.3, and $\Omega_{\Lambda}$=0.7 \citep{Spergel2003}.  We list all transitions by their rounded vacuum wavelengths in the text, but use the unrounded vacuum wavelengths during analysis.

\section{Data Acquisition and Reduction}
\label{observations}

Spectra of SDSS~J0838+2955 were obtained on 2008 February 2 and 6 with the Dual Imaging Spectrograph (DIS) instrument on the Astrophysical Research Consortium (ARC) 3.5 m telescope at Apache Point Observatory (APO). A total of 4.2 hr of spectra of the target were retrieved on the first night of observation and 3.2 hr on the second night.  The 1.5" slit and the highest resolution gratings B1200 and R1200 were used on both nights.  This resulted in a dispersion of 0.62 \AA\ pixel$^{-1}$ and 0.58 \AA\ pixel$^{-1}$ in the blue and red, respectively, with 1.9-2.4 pixels per FWHM resolution element as measured by the calibration lamps.  On the first night, the wavelength centers for the blue and red gratings were 4470\AA\ and 7430\AA\ covering the observed frames of 3850-5095\AA\ and 6825-7990\AA, respectively.  On the second night, the gratings were centered at 4470\AA\ and 5745\AA\ covering the observed frames of 3850-5095\AA\ and 5180-6310\AA.  This resulted in three separate regions: blue, red short, and red long where the blue portion received about double the exposure time of either of the red regions.  The spectra were bias subtracted, flat-field corrected, sky subtracted, wavelength calibrated, extracted, atmosphere corrected with a standard star, and combined using the {\it apall} package in the Imaging and Reduction Analysis Facility \citep{Tody1993}\footnote{IRAF is distributed by the National Optical Astronomy Observatories, which are operated by the Association of Universities for Research in Astronomy, Inc., under cooperative agreement with the National Science Foundation.}.  Considering the variable loss of throughput with guiding errors and cirrus clouds, an eighth degree polynomial was used for each region to flux calibrate the continuum in the ARC spectrum to the SDSS spectrum.   The resulting typical FWHM resolutions were R$\sim$3300, 4500, and 5400 with continuum signal-to-noise ratio (S/N) per FWHM resolution element of $\sim$105, 70, and 80 for each of the three regions (blue, short red, and long red).  Considering the outstanding S/N but rather moderate resolution, the spectrum was binned to one-half the FWHM resolution element for analysis except for the absorption lines from \feii\ which were binned to one full FWHM resolution element to increase the S/N for these shallow troughs.  The systemic redshift of the QSO is 2.043$\pm$0.001 based on the \siIV~$\lambda$1398, \aliii~$\lambda$1859, \feiii~$\lambda$1895, 1914, 1926, and \mgii~$\lambda$2800 emission lines observed in the ARC and SDSS spectra.

\begin{figure}
\centering
\includegraphics[width=4.7in,angle=90]{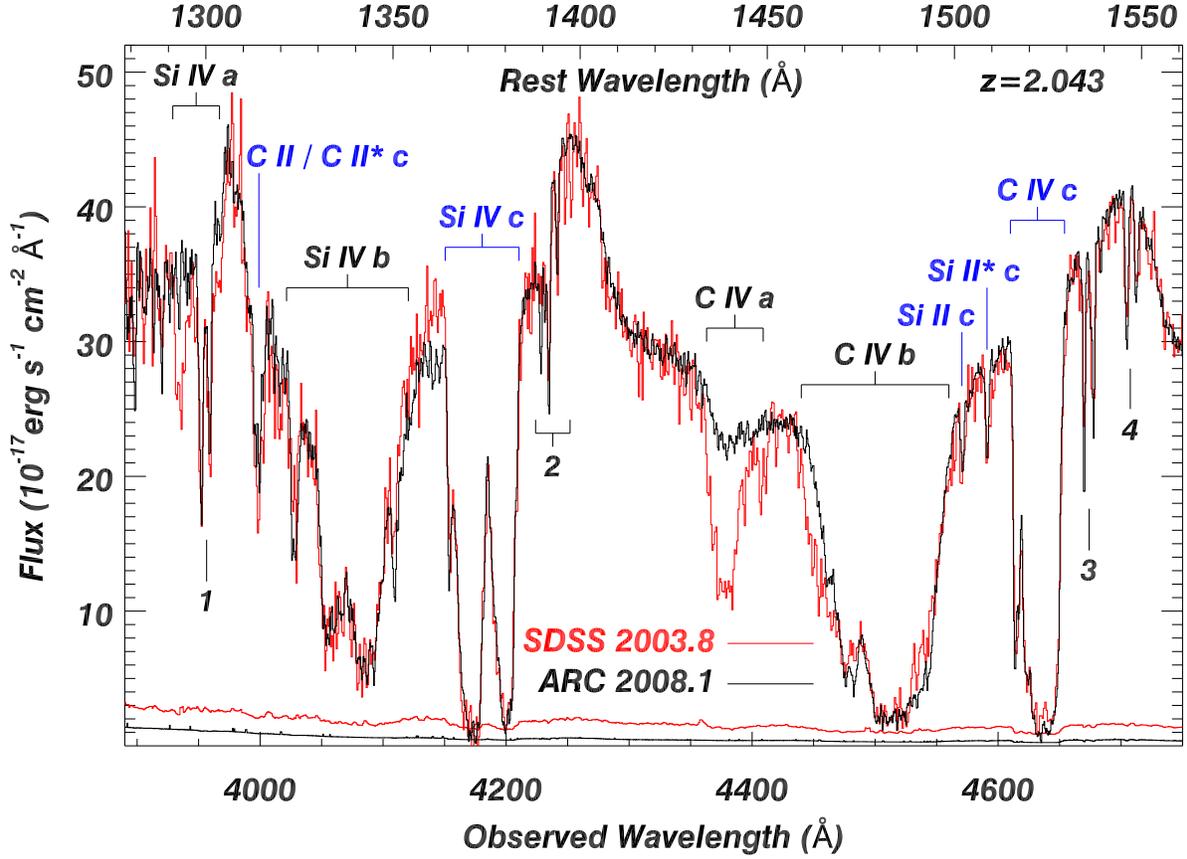}
\caption{Blue portion of the spectrum of SDSS~J0838+2955 comparing the SDSS data (red) taken in late 2003 and the ARC data (black) taken in early 2008.  The three separate outflow components at 22,000, 13,000, and 4900 \kms\ are labeled $a$, $b$, and $c$, while the narrow, most likely intervening systems are labeled 1-4.  Note the variability in component $a$ as well as the high S/N near the \siII\ and \siII* absorption of component $c$ which will allow accurate determination of the number density for that system.  }
\label{BAL.fig}
\end{figure}

We compare the unbinned blue region of the spectra of SDSS~J0838+2955 from both the SDSS and ARC data in Figure~\ref{BAL.fig}.   Note that in this unbinned representation the ARC spectrum has $\sim$3.4 times the S/N and $\sim$1.7 times the spectral resolution compared to the SDSS spectrum. The three separate BAL components at 22,000, 13,000, and 4900 \kms\ are labeled $a$, $b$, and $c$, while the narrow, most likely intervening systems of \civ\ are labeled 1-4 where actually component 2 is a blend of \civ\ intervening and \siIV\ associated with the \civ\ system 4.  Absorption in \civ\ and \siIV\ in outflow component $a$ decreased dramatically between the SDSS spectrum taken in late 2003 and the ARC spectrum taken in early 2008, and this phenomenon is investigated further in Section~\ref{results}.  Note also in the figure the high S/N of the \siII\ and \siII* troughs in component $c$, from which an accurate number density can be determined.

We normalized the continuum by fitting cubic splines across intervals of the spectrum that were unabsorbed.  The systematic errors from normalization are negligible except in the case of the shallow transitions of \feii\ and \feii* where the error is a significant fraction of the absorbed intensity. When calculating column densities for this ion, two additional fits to the continuum were used to estimate the 1$\sigma$ upper and lower bounds on the continuum levels.  We then added in quadrature the errors from normalization to the derived uncertainties of the column densities.

Prior to analysis, some of the data were altered to correct contamination and other reduction errors.  The first instance concerned the \siII*~$\lambda1309$ absorption trough which was contaminated by the red part of an intervening \civ~$\lambda\lambda$1548, 1551 resonance doublet at z=1.528.  Considering the blue component of the intervening \civ\ trough was uncontaminated with an optical depth of $\tau\approx$0.18, its optical depth profile (after dividing by two because of the \civ\ doublet ratio) was used to divide out the contamination of the red component in the \siII* trough.  The second instance was that although the \feii~$\lambda$2383 line appeared centered at the correct location of the other troughs from component $c$, the next two highest strength lines \feii~$\lambda\lambda$2600, 2344 were shifted 0.7 FWHM resolution elements to lower velocities.  Although this much of a shift corresponds to $\sim$6$\sigma$ error based on the wavelength solution, underlying instrumental systematics could have caused such a shift.  The two troughs were shifted back to match with \feii~$\lambda$2383 and therefore the derived column densities for \feii\ should be treated with some caution.

\section{Column and Number Densities for Component $c$}
\label{column}

Ionic column densities were determined for all the transitions covered by the ARC spectrum as well as \mgii\ which was only covered by SDSS.  Atomic data including transition wavelengths, statistical weights, and oscillator strengths are taken from \citet{Tayal2007} for \siII, \citet{Morton2003} for \feii, and \citet{Kurucz1995} for all other ions.  The adopted atomic parameters are displayed in Table 1.

In order to derive the true column densities of the ionic species present in outflow component $c$, one must carefully consider variations in optical depth $\tau$ across the continuum source.  Our group \citep[and references therein]{Arav2005,Arav2008} showed that the apparent optical depth ($\tau = $ -ln [$I$], where $I$ is the residual intensity) method is not a good approximation for outflow troughs (see also below).  For our analysis here, we will consider the covering factor model and inhomogeneous absorber models, and use the fitting routines developed by \citet{Arav2008}.  The covering factor model assumes a constant optical depth across the covered fraction of the continuum source and no absorption across the remainder.  The inhomogeneous model assumes that the optical depth varies as a power-law, $\tau (x) = \tau_{max} x^a$, where the continuum source geometric span $x$ is normalized from zero to unity and the power-law $a$ parameter indicates the extent of inhomogeneity.  

We measured ionic column densities in three different ways to bracket the errors contributed by inhomogeneities of the absorber and saturation of some absorption troughs. Details are given below. Note that all three methods result in rather similar measurements of column densities (see Table 2).  For seven of the ten observed species, the maximum column densities derived were $\ltsim$0.05 dex greater than the lower limits determined by the apparent optical depth method.  For the three optically thick ions of \aliii, \siIV, and \civ, the two different inhomogeneous absorber models were able to bracket the error due to saturation and constrain the column densities to within a factor of 2. The robustness of these extracted column densities lends strength to our photoionization solutions for the outflow (see Section~4), which in turn are used to determine the mass flux and kinetic luminosity of the outflow (see Section~5).

\begin{figure}
\centering
\includegraphics[width=4.2in,angle=90]{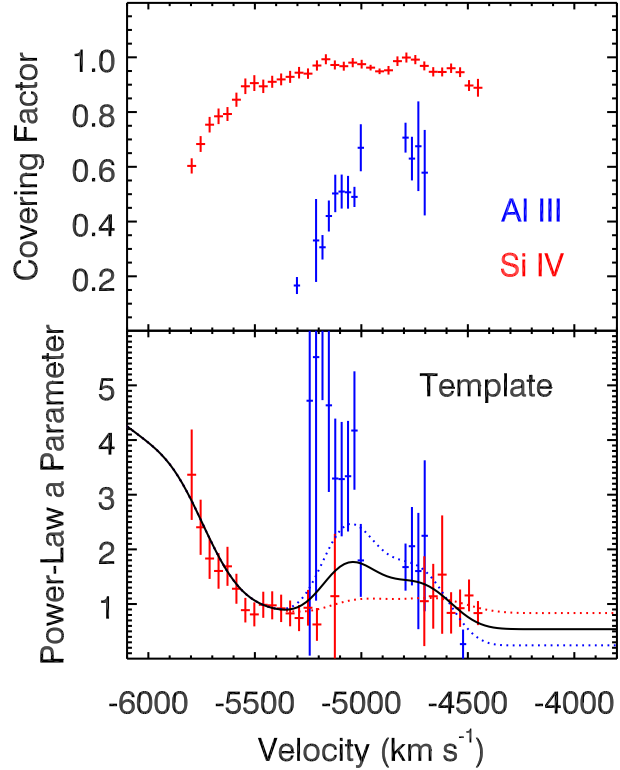}
\caption{Solutions for the covering factor and the power-law $a$ parameter in the inhomogeneous absorber model for \aliii\ (blue) and \siIV\ (red). The covering factor solution (upper panel) is inconsistent between the two ions while the power-law $a$ parameter solutions (lower panel) appear to overlap in the regions where both ions are uncontaminated, e.g. at velocities -4800 to -4400 \kms.  The inhomogeneous absorber model solutions for both ions are smoothed (dotted line), where the high velocity wing of \siIV\ is used to anchor the \aliii\ profile.  The average of these two smoothed solutions is indicated by the template (black), and all three curves will be used to estimate ionic column densities and their uncertainties.}
\label{acov.fig}
\end{figure}

In order to measure deviations from apparent optical depth, one must use two different transitions for the same species to simultaneously fit the ionic column density and the covering factor or inhomogeneous $a$ parameter.  For component $c$, the \siIV~$\lambda\lambda$1394, 1403 and \aliii~$\lambda\lambda$1855, 1863 resonance doublets offer a high enough S/N to constrain both parameters; however, \siIV\ is saturated near the velocity center and \aliii\  is contaminated from atmospheric emission of \oi~$\lambda$5577 at the center (-4900 \kms) of the red component and by an intervening absorption system at z=1.333 of \feii~$\lambda$2383 in the low-velocity wing (-4600 \kms) of the blue component.  Nevertheless, we extracted solutions to the covering factor model and power-law $a$ parameter in the inhomogeneous model for the velocity bins without contamination or blending, and these are shown in Figure~\ref{acov.fig}.  Note that the solutions to the covering factor are quite discrepant between the two different species of \aliii\ and \siIV, while the inhomogeneous model appears to be consistent between the two ions over the few velocity bins for which both ions are non-contaminated.  We constructed a model for the power-law $a$ parameter by smoothing the solutions with a moving boxcar average of nine velocity bins (dotted lines in figure).  Considering the large error in the $a$ parameter solution for \aliii\ within the high velocity wing, we used the \siIV\ $a$ profile between -5900 and -5250 \kms\ to anchor the smoothed \aliii\ profile.  Although the smoothed $a$ profile for \aliii\ is significantly lower than its actual solution at velocities -5200 to -5000 \kms, the derived total \aliii\ column densities differ by only $\sim$6\%.  The average of the two smoothed $a$ profiles is indicated by the black template line in the figure.  This average template solution is used for reporting minimum $\chi^2_{\nu}$ values as well as ionic column densities in the figures and text, except for \aliii\ and \siIV\ in which case their own respective smoothed solutions (dotted lines in figure) have been used.  The average template solution is not used in Table 2 (unless otherwise stated), but instead the smoothed solutions for \aliii\ and \siIV\ are implemented to bracket the errors.  

\begin{figure}
\centering
\includegraphics[width=7.0in,viewport=80 20 430 360,clip]{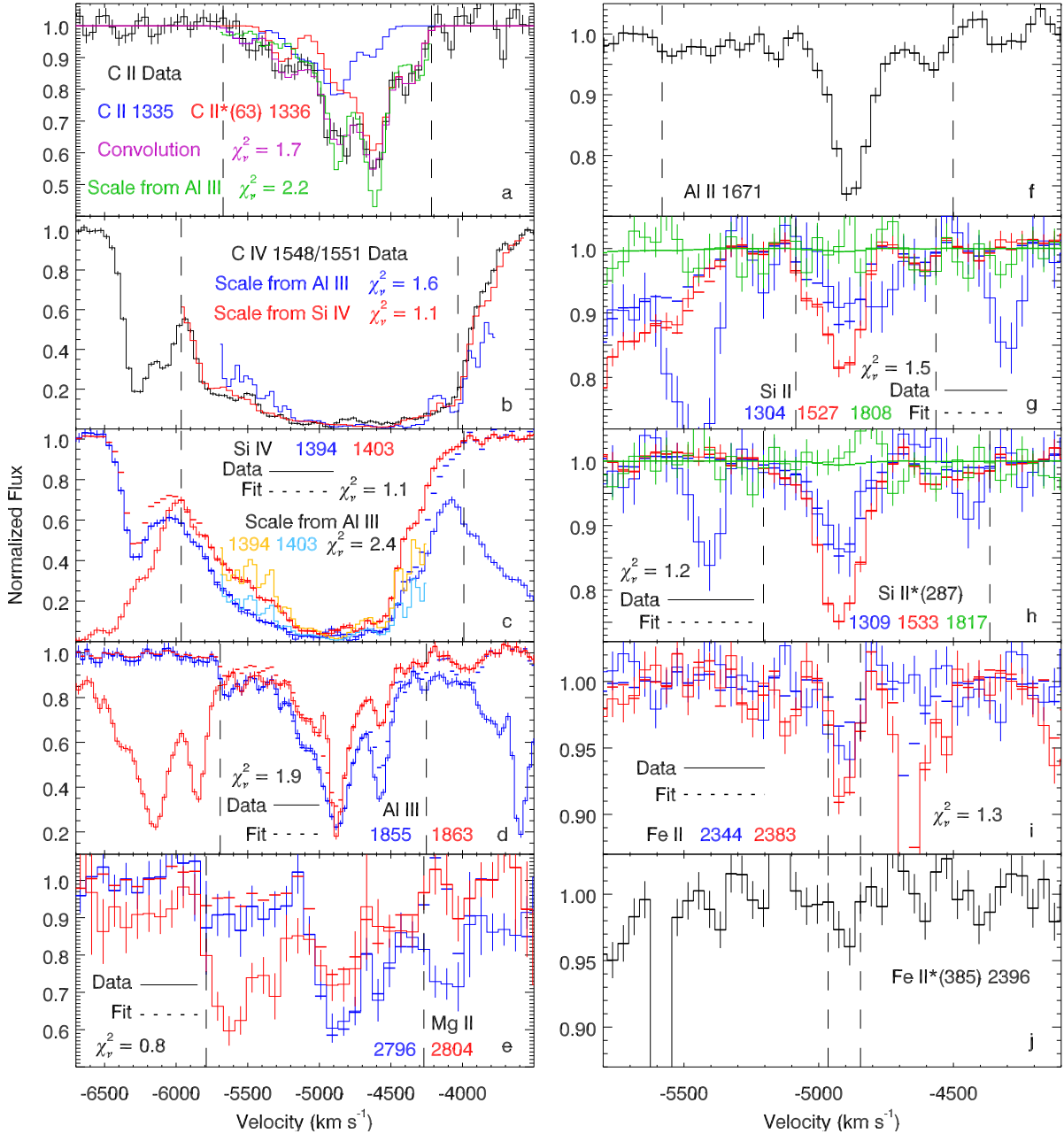}
\caption{Data and fit to the absorption transitions observed in outflow component $c$.  For each panel, the ion and wavelengths in \AA\ of transitions are given, and the reduced $\chi^2$ value is indicated if the $\chi^2$ statistic was minimized.  The adopted inhomogeneous absorber model is used to fit each velocity bin for the specific species.  Gaussian fittings and scaling methods from other ions were employed for \cii, \civ, and \siIV\ (panels (a)-(c)), which all gave consistent results. (See text for full details.)}
\label{fit.fig}
\end{figure}

We show all of the absorption features of component $c$ as well as the fits to the data in Figure~\ref{fit.fig}. We integrated all column densities over the full profile of the trough (between the dashed lines in the figures), up to a maximum range between -6000 and -3900 \kms. The high velocity component at -6250 \kms\ is excluded because it is kinematically separated, seen only in the higher ionization species of \civ\ and \siIV\ and is in any case negligible, representing less than 2\% the total true ionic column densities.  Although we use only one power-law $a$ parameter solution to report the column densities in the text below, we use two profiles to estimate the error in the column densities contributed by the uncertainty in the power-law $a$ solution.  For \siIV\ and \aliii\ we used their own smoothed solutions (dotted red and blue lines in Figure~\ref{acov.fig}, respectively) and the average template solution (black line).  For all other ions, we compare the \siIV\ and \aliii\ smoothed solutions to determine the error in N$_{ion}$ due to the uncertainty in the power-law $a$ parameter.  The measured column densities assuming the apparent optical depth model, the \siIV\ power-law $a$ model, the \aliii\ power-law $a$ model, and the adopted solution are presented in Table 2.  The adopted ionic column densities are derived such that they span the values and 1$\sigma$ errors of the two power-law $a$ models. The details of ionic column density extractions are given henceforth.

The ground transition of \cii~$\lambda$1334.53 is blended with the transitions from its excited state of E=63.4 cm$^{-1}$ at $\lambda\lambda$1335.66, 1335.71 (panel (a)).  Both transitions of the excited state were combined into a single transition where the sum of the oscillator strengths and weighted wavelength transition were used. Since the electron number density log $n_e$ (cm$^{-3}) \approx$ 3.7 based on the \siII*/\siII\ ratio (see below) is above the critical density of log $n_e$ (cm$^{-3}) \approx$ 2.0 for \cii* (E=63 cm$^{-1})$, the level population ratio of $N$(ground)/$N$(excited) is nearly the Boltzmann ratio which is essentially the ratio of statistical weights (1:2) for T $\sim$ 10$^4$ K.  Because the ground and excited states of \cii\ are blended, the column density cannot be measured directly, and therefore a profile for the optical depth must be modeled. The blended profile was deconvolved by fitting Gaussian optical depth profiles to the ground and excited states assuming an inhomogeneous absorber model defined by the template power-law $a$ parameter profile determined above (blue, red, and purple lines in panel (a)).  Four pairs of Gaussians were used for the ground and excited state profiles, each pair with three free parameters (line center, optical depth, and Doppler b parameter). Minimization of the $\chi^2$ statistic resulted in a reduced $\chi^2$ of 1.7 and a column density of $(8.2\pm0.3) \times 10^{14}$ cm$^{-2}$. 

A second model for the column density profile of \cii\ was assumed to be the same as the solution for \aliii\ (green line in panel (a)). This is motivated since \cii\ and \aliii\ have similar ionization potentials and therefore the ratio of column densities between \cii\ and \aliii\ should be nearly constant.  We multiplied the \aliii\ column density profile by a single scaling factor over all velocity bins and then convolved with the power-law $a$ parameter to produce the absorption profile.  Two separate absorption profiles were convolved for each the ground and excited state, again assuming their ratio is 1:2, to produce the total absorption profile for the entire trough. When fitting, the error for each velocity bin is taken to be the error of the data for \cii\ added in quadrature with the error in the column density solution of \aliii. This additional error from the \aliii\ column density solution is why the fit appears significantly worse yet with only a slightly larger reduced $\chi^2$ value of 2.2. This scaling method resulted in a column density of $(8.8\pm0.7) \times 10^{14}$ cm$^{-2}$, which is only 0.03 dex higher than the solution found from the Gaussian deconvolution solution described previously. Both the systematic errors that arise from the different methods employed as well as the uncertainty in the $a$ profile are accounted when reporting the adopted column density solution in Table 2.

The transition of \civ~$\lambda\lambda$1548, 1551 is broad, blended and deep with residual intensities of only $\sim$2\% (panel (b)). Considering that the central portions of the \civ\ profile are saturated, the scaling of optical depth method was done with \aliii\ and \siIV\ as it was done for \cii. We also convolved the scaled model troughs of the blue and red components of the \civ\ doublet when fitting the entire absorption profile.  This is why the red component of the trough extends the doublet separation of $\sim$500 \kms\ to the right of the dashed line in the figure.  The two different scaling fits from \aliii\ and \siIV\ resulted in consistent column densities of $(2.07\pm0.27) \times 10^{16}$ cm$^{-2}$ and $(1.67\pm0.19) \times 10^{16}$ cm$^{-2}$ with a reduced $\chi^2$ of 1.6 and 1.1, respectively.  

The transitions of \siiv~$\lambda\lambda$1394, 1403 are well resolved but nearly saturated at the center (panel (c)). The fit using the smoothed \siIV\ power-law $a$ parameter profile resulted in a column density of $(8.4\pm0.5) \times 10^{15}$ cm$^{-2}$.  The scaling method from the solution for \aliii\ was also used, but in this case both the blue and red troughs of the doublet are fitted simultaneously as they are not blended and so there is no need for convolution; the light blue line is the scaled fit to the blue component of \siIV\ and the orange line is the scaled fit to the red component.  The resulting column density of $(1.12\pm0.17) \times 10^{16}$ cm$^{-2}$ is only 0.12 dex higher than measuring \siIV\ on its own merits.

The reasonable $\chi^2$ fits when scaling the optical depth profile of one ion to a different ion as well as the consistency in measuring column densities using the various techniques suggest that the ratio of column densities, even between species with different degrees of ionization, is fairly constant across the velocity profile.  This is further evidenced by the high velocity wing between -5600 and -5200 \kms\ that appear as shallow absorption features in the low-ionization species of \aliii, \mgii, and \alii, but as deep troughs in the high ionization species of \civ\ and \siIV.  Although the \civ\ and \siIV\ absorption features are $\sim$2000 \kms\ wide, most of the troughs are low-column density wings while the majority of the column density is located at the central $\sim$200 \kms\ of the profile.

For the remaining six species, \aliii, \mgii, \alii, \siII, \siII* (E=287 cm$^{-1}$), \feii, and \feii* (E=385 cm$^{-1}$; panels (d)-(j)), we used the profile of the power-law $a$ parameter to fit the column densities for each velocity bin.  Although we plot only the two transitions $\lambda\lambda$2344, 2383 for \feii\ in panel (i), we used two additional lines $\lambda\lambda$2587, 2600 in the fit. The measured column densities assuming apparent optical depth and the various inhomogeneous absorber models are presented in Table 2. Note that the apparent optical depth method underestimates the true column densities by 0.2-0.8 dex for the optically thick, higher ionization species of \aliii, \siIV, and \civ.  This failure of the apparent optical depth method has been observed previously in a variety of AGN exhibiting outflows \citep{Arav1997, Hamann1997, Arav1999a, Arav1999b, Arav2001, Arav2002, Scott2004, Gabel2005}.

\begin{figure}
\centering
\includegraphics[width=2.6in,angle=90]{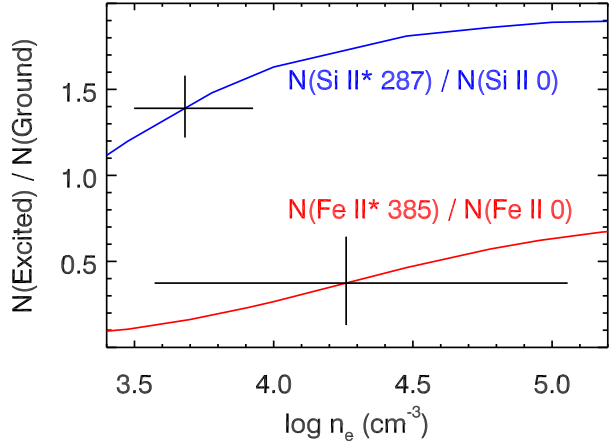}
\caption{Theoretical density-dependent ratio of column densities of \siII\ and \siII* ($E = $287 cm$^{-1}$) as well as of \feii\ and \feii* ($E = $385 cm$^{-1}$) for $T_e$ = 10$^4$ K.  The uncertainty in the density is indicated based on the measured error of the ratios of column densities.  }
\label{density.fig}
\end{figure}

To determine the electron number density of the absorber, $n_e$, we use the column densities of the ground and first excited levels of \siII, 3s$^2$3p~$^2$P$^o_{1/2}$ and 3s$^2$3p~$^2$P$^o_{3/2}$ at 287~cm$^{-1}$ respectively, and of \feii, $a~^6$D$_{9/2}$ and $a~^6$D$_{7/2}$ at 385 cm$^{-1}$ respectively.  To ensure consistency when calculating number densities, the measured column densities are integrated over the same velocity region for each species instead of using the total column densities reported in Table~2. To determine the density dependent ratio of column densities, we use the 142-level spectral model of \feii\ of \citet{Bautista1998} and a \siII\ model built with atomic data from the CHIANTI v.~5.2 atomic database \citep{Landi2006}, both assuming an electron temperature $T_e \approx$10$^4$ K.  The theoretical ratio of column densities are plotted in Figure~\ref{density.fig}, and the measured ratio in column densities are indicated by the vertical lines.  The derived density is simply the range in the ratio curve that spans the measured ratio.  Both line ratios yield consistent diagnostics and we adopt the weighted average of log $n_e$ (cm$^{-3}$) = 3.75 $\pm$ 0.22, where the uncertainty is entirely from measurement statistics.  

With the number density now determined, the total column density in \siII\ is $N$(total) = $N$(ground) + $N$(E=287 cm$^{-1}$). To determine that of \feii\ we rely on our model \feii\ atom as well as the above electron density and its uncertainty to find that $N$(total) = (2.5 $\pm$ 0.3)$ \times N$(ground). These measured column densities (as well as the photoionization model predictions - see below) are given in Tables 3 and 4.

\section{Photoionization Analysis}
\label{photoion}

\subsection{Spectral Energy Distribution}

\begin{figure}
\centering
\includegraphics[width=2.6in,angle=90]{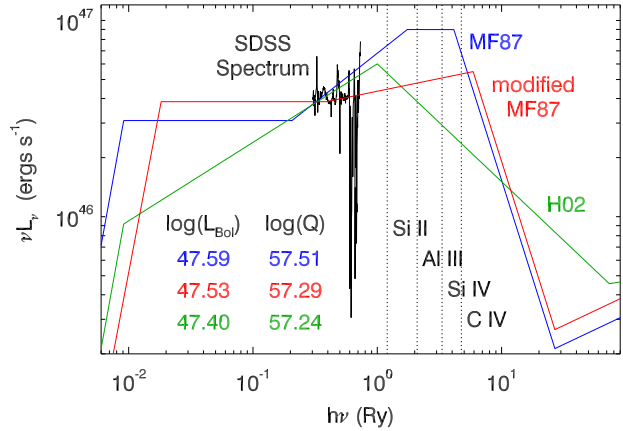}
\caption{Three adopted SEDs normalized to the SDSS spectrum.  The bolometric luminosity and rate of ionizing photons $Q$ are calculated.  The ionization potentials for destruction are indicated by vertical lines for various ions.}
\label{sed.fig}
\end{figure}

The photon (SED) incident on the outflow is important in determining the ionization and thermal structures of the absorbing plasma. Because the ions used in photoionization modeling have ionization potentials for destruction in the far ultraviolet (FUV), it is important to accurately model this region of the SED.  We first consider the standard \citet[][hereafter MF87]{Mathews1987} SED as well as the SED used in \citet[][hereafter H02]{Hamann2002}.  After normalizing these SEDs near 2500 \AA\ in the rest frame of SDSS~J0838+2955 to the flux calibrated SDSS spectrum (see Figure~\ref{sed.fig}), it becomes apparent that the observed SED is softer in the UV than either of these two SEDs.  The slope beyond 0.4 Ryd was found to be $\alpha \approx$ -0.87 (F$_{\nu} \propto \nu^{\alpha}$), about 0.3 dex softer than the MF87 or H02 SED.  This discrepancy can be due to either an inherent difference in the intrinsic quasar continuum or extinction due to dust grains either within the outflowing gas or somewhere within the host galaxy. In the latter case, grain extinction was found to diminish the rate of hydrogen ionizing photons by only $\Delta $log $Q = 0.18\pm$0.09 and produce a color excess of E(B-V)$\approx$0.03 mag, using the method described in \citet{Dunn2009} that relies on relatively unobscured Two Micron All Sky Survey (2MASS) photometry data in the near IR. Considering that we do not know whether the deviation from the theoretical SED is due to an inherent difference in the SED, dust extinction that occurs within the absorber, or extinction that occurs close to the AGN, we simply modified the MF87 SED to match the observed UV continuum (also shown in Figure~\ref{sed.fig}).  In addition to matching the observed slope in the UV, the UV-optical break was reset to 0.4 Ryd to better coincide with the SDSS spectrum, and the IR break was reset to 5$\mu$m to produce a more realistic heating source. We computed the bolometric luminosities and rate of ionizing photons $Q$ for each of the three SEDs, and the scatter in both quantities are $\ltsim$0.2 dex. We consider it plausible that the actual incident FUV SED falls somewhere between our three adopted models.

\subsection{Photoionization Modeling}

We now examine the formation of the various ionic species observed in our spectrum.  For the present work, we use the photoionization modeling code Cloudy \citep{Ferland1998} v08.00 to compute spectral models of the outflowing absorbing gas in SDSS~J0838+2955. We start by assuming constant total hydrogen density ``clouds'' using Cloudy's default solar abundances in a plane-parallel geometry. In building a fully self-consistent photoionization model, we adopt a hydrogen number density $n_H$ that is 0.05 dex lower than the electron number density, i.e  log $n_H$~(cm$^{-3}$) = 3.70$\pm$0.22. This assumption is motivated by the photoionization solutions (see below) which predict the thickness of the cloud to be 0.2 -- 0.3 dex less than the column density necessary to reach the hydrogen ionization front.  Considering the cloud contains mostly ionized hydrogen and helium, the electron number density in the gas region where \siII\ and \feii\ are formed is $\sim$1.1 times higher than the hydrogen number density.  Moreover, small variation of $n_H$ have negligible effects on the results of photoionization modeling. We used Cloudy's default 16-level population for \feii, because for this low gas density of log $n_e$ (cm$^{-3}$) $\approx$ 3.75 at T $\approx$ 10$^4$ K, increasing the number of population levels had a negligible effect on the column densities.

We constructed a grid of models covering the parameter space of total hydrogen column density $N_H$ and ionization parameter $U_H$.  Here, the ionization parameter is defined as $U_H \equiv \Phi_H/c {n_H}$, where $\Phi_H$ (cm$^{-2}$~s$^{-1}$) is the hydrogen ionizing photon flux.  The grids spanned log~$N_H$ (cm$^{-2}$)  = 20.20 - 21.50 and log~$U_H$ = -2.40 to -1.40, with 0.05 dex steps for both parameters.  The 27$\times$21=567 individual models for a single grid were computed for solar abundances only.  To determine the effects of varying the abundances of silicon and carbon, we scaled the column densities of these species from the solar abundance grid output and then compared this scaled grid model to the observed column densities. This approach is justifiable because small variations in the silicon and carbon abundances have negligible effects on the ionization structure of the cloud.  Thus, additional model computations are unnecessary.

When comparing the ionic column densities predicted by the models to the observed column densities, we used two different methods to constrain $N_H$ and $U_H$. The first approach entails using pairs of ionic column densities from a single element, specifically \cii\ and \civ\ or \siII\ and \siIV, to determine a solution to $N_H$ and $U_H$. The strength of this method is that it is relatively immune to elemental abundance ratios differing from solar.  Solutions to the \alii\ and \aliii\ column density pair are not considered because there is an apparent problem in modeling the column densities of these two ions as measured in AGN outflows \citep{Korista2008, Dunn2009}.

The second approach in measuring $N_H$ and $U_H$ is to compute the $\chi^2$ statistic for the measured ionic column densities, similar to what was done in \citet{Arav2007}.  We define the reduced $\chi^2$ as

\begin{equation}
\chi^2_{\nu} = {{1} \over {\nu}} \sum_{i} \left({{\log N_{i,model} - \log N_{i,observed}} \over {\sigma_{i,dex}}} \right)^2
\end{equation}

\noindent where $i$ represents the index of one of the ionic species measured in the spectrum, $\sigma_{i,dex}$ is the measured error in dex of the ionic column densities, and $\nu$ is the number of degrees of freedom.  When minimizing $\chi^2_{\nu}$, as well as determining the solution using a pair of ions, the grid of models is interpolated using a cubic convolution technique so that the true solution may be estimated.

\begin{figure}
\centering
\includegraphics[width=3.8in]{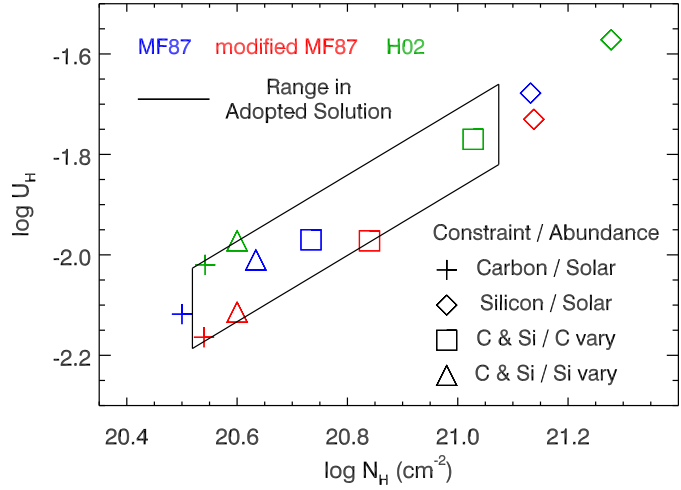}
\caption{Parameter space of total hydrogen column density $N_H$ and ionization parameter $U_H$.  The three colors represent the different SEDs used: MF87 (blue), modified MF87 (red), and H02 (green). The different symbols are the solutions to $N_H$ and $U_H$ using different constraints and chemical abundances.  The $+$'s are the solutions to $N_H$ and $U_H$ using only the \cii\ and \civ\ column densities assuming solar abundances.  The $\Diamond$'s are the solutions using only \siII\ and \siIV\ column densities assuming solar abundances.  The $\Box$'s are the solutions to the $\chi^2$ statistic using the four ions \cii, \civ, \siII, and \siIV\ as constraints, but where the abundance of carbon was allowed to decrease. The $\triangle$'s are the solutions to the $\chi^2$ statistic where the abundance of silicon was allowed to increase.}
\label{photoionfit.fig}
\end{figure}

The solutions to $N_H$ and $U_H$ are shown in Figure~\ref{photoionfit.fig} where the different colors represent the different SEDs and the various symbols represent the solutions using different constraints.   For all SEDs, the solutions using the \cii\ and \civ\ column densities ($+$'s) are $\sim$0.6 dex smaller in $N_H$ compared to the solutions using the \siII\ and \siIV\ column densities ($\Diamond$'s). To study the effect of altering abundances, the $\chi^2_{\nu}$ statistic is minimized with the ability to vary the carbon abundance ($\Box$'s) or the silicon abundance ($\triangle$'s). When computing $\chi^2_{\nu}$, only four ions are used: \cii, \civ, \siII, and \siIV.  Considering one must constrain both $N_H$ and $U_H$ as well as the chemical abundance of the single element allowed to vary, then the number of degrees of freedom is $\nu$ = 1. We present the predicted model parameters for these solutions in Tables 3 and 4.  Most of the solutions that use the minimization technique result in $\chi^2_{\nu}$ values $\sim$1, suggesting the fit to $N_H$ and $U_H$ is rather good.

Depending on the model, the (Si/C) abundance ratio increased 0.1 -- 0.6 dex relative to the solar ratio, as is expected in gas of super-solar metallicities \citep{Ballero2008}. Adopting a reasonable confidence interval that the (Si/C) abundance ratio is 0.2 -- 0.5 dex greater than solar, then the metallicity of component $c$ is Z/\Zsun = 1.6 -- 3.1 based on the interpolation of the chemical abundances in \citet[][their Table 2]{Ballero2008}.  This super-solar metallicity is also consistent with the chemical abundances determined for the outflow in Mrk 279 \citep{Arav2007}.  Even if the metallicity were as high as Z/\Zsun$\approx$3, the carbon abundance would increase by $\approx$0.1 dex relative to solar \citep{Ballero2008} and therefore the derived solution to $N_H$ would only be 0.1 dex less than the solution based on the carbon ions.

An important conclusion drawn from the above models is that similar estimates for $N_H$ and $U_H$ are obtained despite the wide range of options in the adopted SED and chemical composition of the cloud. Taking the averages and standard deviations of all twelve solutions results in log~$N_H$ (cm$^{-2}$) = 20.80 $\pm$ 0.28 and log~$U_H$ = -1.93 $\pm$ 0.21.  However, rather than assuming $U_H$ to be independent from $N_H$, we determine a least-squares linear fit to the twelve solutions and adopt the correlation function to be

\begin{equation}
 \mathrm{log}~U_H = -1.93 + 0.6594 \times (\mathrm{log}~N_H / 10^{20.8} \mathrm{cm}^{-2}) \pm 0.08,
\end{equation}

\noindent which is displayed as the black parallelogram in the figure.  This treatment of $U_H$ and $N_H$ will significantly reduce the errors in the derived mass of the outflow, and hence better constrain the mass flux and kinetic luminosity (see below). 

Figure~\ref{cloud.fig} shows a representative photoionized single-slab model using the MF87 SED, log $U_H$ = -1.95 and chemical abundances where carbon is decreased by 0.3 dex below solar while silicon is increased by 0.2 dex. We plot the ionization fraction for the species apparent in absorption and indicate with squares the hydrogen column densities necessary so that the model ionic column densities matches the observed ionic column densities for the given $U_H$. The adopted value and error in $N_H$ are displayed as the vertical lines.  Note that all the displayed squares are within 0.2 dex of the estimated solution for $N_H$.  Note also that the ionic fractions of \cii, \mgii, \siII, and \feii\ rise rapidly approaching the hydrogen ionization front at log~$N_H$ (cm$^{-2}$) $\approx$ 21.05 in this model (see \citealt{Korista2008} for discussion).  The expected column density of the absorber is 0.2 -- 0.3 dex less than this necessary column to reach the ionization front, and therefore the lower ionization species are fairly insensitive to constraining $N_H$ in this regime.    

The only ion that appears to be considerably different than the observed is \feii, where the models predict the \feii\ column density should be 10 times less than measured (see Tables 3 and 4).  Considering the photoionization solutions were determined in great part by the silicon column densities, both the iron abundance (Fe/H) and the (Fe/Si) abundance ratio would have to be a full order of magnitude greater than solar to explain the discrepancy in the \feii\ column densities.  However, such an increase in the (Fe/Si) abundance ratio is not predicted by chemical evolution models \citep{Ballero2008}.  

In the representative slab model, the electron temperature and electron number density are fairly constant across the slab with deviations from uniformity that are $\ltsim$0.2 dex and $\ltsim$0.05 dex, respectively.  The ionic electron temperatures of \siII\ and \feii\ were 10,400 $\pm$ 500 in the model, justifying our assumption of $T_e \approx$10$^4$ K when we derived the electron number density from the level populations of \siII\ and \feii.

In order to compute the distance of component $c$ to the ionizing source, we must determine whether the flux of ionizing photons arriving to the absorber may have suffered from attenuation from intervening gas. Our spectrum does show strong \siIV\ and \civ\ absorption from a higher velocity system (component $b$), which might arise interior to component $c$. No absorption from lower ionization species can be measured on their own merits, suggesting this system is more ionized and/or lower in hydrogen column density than component $c$. In order to determine the attenuation effect, we estimated from the spectrum a 2$\sigma$ upper limit to the \cii\ column density of $9.6 \times10^{13}$~cm$^{-2}$ in component $b$ together with a lower limit to N(\civ) of $8.8\times10^{15}$~cm$^{-2}$ from the apparent optical depth of the \civ\ trough of component $b$. These limits yield a lower bound of log~$U_H > $ -1.95 and an upper bound  of log $N_H$ (cm$^{-2}$) $<$ 20.4 (cm$^{-2}$). The attenuation of ionizing photons through such a cloud with these parameters is bounded by $\Delta$log $Q <$ 0.07 dex.  Thus, the uncertainty in the chosen SED as well as the range of solutions in $U_H$ and $N_H$ for the system of interest (component $c$) dominates the error compared to the uncertainty of furthered attenuation by component $b$.  We therefore ignore component $b$ and assume it has no effect on the photoionization solution to component $c$.

\section{Results and Discussion}
\label{results}

With estimates of $n_H$, $N_H$, $U_H$, and $Q$ in hand, the large-scale physical properties of the outflow can be calculated.  From the definition of the ionization parameter \citep[][Equation (14.7)]{Osterbrock2006}, the distance of the outflow from the central AGN is

\begin{equation}
R = \left( {{Q} \over {4 \pi c U_H n_H}}\right)^{{1} \over {2}} = 3.3^{+1.5}_{-1.0}~ \mathrm{ kpc}.
\end{equation}

\noindent This large distance with a velocity v=4900 \kms\ suggests that this outflow will continue to expand into the IGM, where its mass and kinetic energy will be dissipated as heat into the intracluster medium.  Assuming the outflow is in the form of a thin partial shell, its mass is

\begin{equation}
M = 4 \pi \mu m_p \Omega R^2 N_H = (1.9^{+1.8}_{-0.9}) \times 10^8~\Omega_{0.2}~ \mathrm{\Msun}
\end{equation}

\noindent where $\mu \approx$ 1.43 is the mean atomic mass for solar abundances, $m_p$ is the mass of the proton, and $\Omega$ is the portion of the solid angle subtended by the shell. The adopted parameter space for $N_H$ and $U_H$ is used to determine the expected uncertainty in the mass.  A full discussion about what $\Omega$ value is appropriate for these outflows is given in \citet{Dunn2009}.  To summarize their argument: there are good reasons to assume that the $\Omega$ of these low-ionization BAL outflows with excited metastable states is similar to the canonical value for \civ\ BALs \citep[$\Omega \sim$ 0.2; see][]{Hewett2003}, and that their scarcity is due to the selection effect that low-ionization BALs are rare in general. See also \citet{Hall2003} for further discussion regarding this point.  In order to accommodate for the uncertainty in $\Omega$, we leave the final answer as a function of $\Omega_{0.2} \equiv \Omega / 0.2$. The mass flux is then given by

\begin{equation}
\dot{M} = 8 \pi \mu m_p \Omega R N_H v = 590^{+410}_{-240}~ \Omega_{0.2}~ \mathrm{\Msun\ yr}^{-1} 
\end{equation}

\noindent where the adopted parameter space for $N_H$ and $U_H$ is used.  If the solution to $N_H$ and $U_H$ were assumed to be uncorrelated, then the error in  $\dot{M}$ due to photoionization modeling would be 0.30 dex instead of the actual 0.19 dex.  Finally, the kinetic luminosity of the outflow is

\begin{equation}
\dot{E}_k = \frac{\dot{M} v^2}{2} = 4 \pi \mu m_p \Omega R N_H v^3 = (4.5^{+3.1}_{-1.8}) \times 10^{45}~ \Omega_{0.2}~ \mathrm{erg~s}^{-1}
\end{equation} 

\noindent which is (1.4$^{+1.1}_{-0.6}$ $\Omega_{0.2}$)\% the bolometric luminosity of the QSO itself.  The kinetic luminosity of this component is sufficiently large to affect its host environment and contribute to AGN feedback mechanisms.  Indeed, this value of $\sim$1.4\% is more likely a lower limit considering that the other outflow components $a$ and $b$ could further contribute to AGN feedback \citep[see][]{Arav2009}. Moreover, in Seyfert galaxies the hot phase of the outflow seen in X-rays can carry 70\%-99\% of the kinetic luminosity of the outflow \citep{Gabel2005b,Arav2007}.  We point out that a few earlier investigations have found outflows with similar physical properties.  Based on the \citet{Hamann2001} analysis of QSO 3C~191,  the kinetic luminosity of the outflow observed in this object is $\sim$9$ \times 10^{43}~ \Omega_{0.2}$ erg s$^{-1}$ and its mass flux is $\sim$310$~ \Omega_{0.2}$ \Msun\ yr$^{-1}$  for an outflow situated at $\sim$28 kpc.  In QSO~FBQS~1044+3656, \citet{deKool2001} found the outflow to be $\sim$1 kpc from the AGN with a kinetic luminosity of $\sim$9$ \times 10^{44}~ \Omega_{0.2}$ erg s$^{-1}$ and mass flux of $\sim$150$~ \Omega_{0.2}$ \Msun\ yr$^{-1}$.

Given the bolometric luminosity and assuming that the black hole is accreting near the Eddington limit, we find that the amount of mass being accreted is \citep{Salpeter1964}:

\begin{equation}
\dot{M}_{acc} = {{L_{Bol}} \over {\epsilon_r c^2}} \approx 60~\mathrm{\Msun\ yr}^{-1} 
\end{equation}

\noindent assuming an accretion efficiency $\epsilon_r$ of 0.1. The mass flux of the outflow is $\sim$10 times greater than the accretion rate, which is physically plausible assuming the outflow entrained a substantial amount of material on its journey to $\sim$3 kpc away from the AGN. 

Furthermore, SDSS~J0838+2955 is an example of a QSO with outflows at different distance scales considering that the outflow contributing to component $a$ is most likely much closer than component $c$.  The argument is as follows.  The significant variation in optical depth of component $a$ is most likely due to a change in flux of the ionizing source or transverse velocity of the outflow out of the line of site to the central AGN \citep{Barlow1994}.  Although the spectrum itself cannot be used to determine variations in luminosity (because the ARC spectrum was flux calibrated to the SDSS spectrum), several slit guider images were used to conduct differential photometry between the target and foreground stars of similar color in the field.  The change in $r$ magnitude (the color corresponding to the peak efficiency of the guider CCD) was determined to be less than 0.09 mag at the 2$\sigma$ level and thus the variation in absorption of the outflow was most likely due to transverse motion.  

An upper limit on the distance of this component $a$ of the outflow can be estimated assuming that the transverse velocity cannot exceed the Keplerian velocity.  The Keplerian velocity is determined from the mass of the black hole $M_{BH}$ which in turn is estimated from a mass-luminosity relation.   In more detail, the luminosity of the QSO is expected to be $\nu L_{\nu} \approx 3 \times 10^{46}$ erg s$^{-1}$ near the restframe wavelength of 5100 \AA.  From Equation (9) in \citet{Peterson2004} and assuming L/L$_{Eddington} \approx$0.1, the mass of the central black hole is expected to be $M_{BH} \approx 7 \times 10^{9}$ \Msun, although the mass--luminosity relation is not well constrained in this high-mass regime.  From this black hole mass, the radius of the UV continuum region arising from the accretion disk can be estimated to be $\sim$30 light-days \citep[Equation (3.20) in][]{Peterson1997} assuming the accretion rate is near the Eddington limit.  We will choose a lower limit on the line of sight covering factor of component $a$ to be $\sim$0.5 because the \civ\ absorption trough had an initial residual intensity of $\sim$0.43. The transverse velocity of the outflow across half of the $\sim$60 light-day wide continuum source in the restframe time interval of 4.3 yrs / 1+z = 1.4 yrs must be v $\gtrsim$ 30 light-days / 1.4 yrs $\approx$ 18,000 \kms\ in the restframe of the QSO.  Finally, assuming the tangential velocity of the outflow does not exceed the Keplerian velocity, then the distance of outflow $a$ from the central AGN must be R $<$ $G M_{BH}/v^2 \approx$  0.1 pc.  

\section*{Conclusions}

1. Outflow component $c$ of SDSS~J0838+2955 has a velocity of 4900 \kms\ with respect to the rest frame of the QSO and exhibits absorption from ground and excited metastable states of \siII\ and \feii.  From the ratios of the column densities of the excited to ground states, the number density of the outflow was determined to be log $n_e$ (cm$^{-3}$) = 3.75 $\pm$ 0.22.

2. The solutions to the \aliii\ and \siIV\ absorption profiles of component $c$ indicate inhomogeneity of the absorber across the continuum source.  The apparent optical depth method underestimates the true column density by 0.2-0.8 dex for the optically thick species of \aliii, \siIV, and \civ.  

3. The uncertainty in the incident SED and chemical abundances of the outflow resulted in only slight variations in the derived ionization structure of the cloud.  The total hydrogen column density for component $c$ was determined to be log $N_H$ (cm$^{-2}$) = 20.80 $\pm$ 0.28, and the ionization parameter was found to be log $U_H$ = -1.93 $\pm$ 0.21.  The derived column density is 0.2 -- 0.3 dex smaller than the column density necessary to reach the hydrogen ionization front.  

4. There is a strong indication that outflow component $c$ has super-solar metallicities, ranging in Z/\Zsun = 1.6--3.1, consistent with other reliable abundance estimates of AGN outflows \citep{Gabel2006,Arav2007}.

5. We measured the distance of this outflow to be 2.3-4.8 kpc with a kinetic luminosity of 0.8\%--2.5\% the bolometric luminosity of the QSO. This large kinetic luminosity suggests that QSO outflows can serve as a major contributor to AGN feedback.

\acknowledgments

We acknowledge support from NSF grant AST 0507772. M.M. and N.A. thank the Astrophysical and Planetary Sciences department at the University of Colorado for the use of the ARC 3.5 m telescope. M.M. also gratefully acknowledges the Center for Astrophysics and Space Astronomy for the use of their computing facilities, as well as enlightening discussions with John Stocke.    

\bibliographystyle{apj}
\bibliography{biblio}

\begin{deluxetable}{lccccc}
\label{atomic}
\tablecaption{Atomic Data for the Transitions Used}
\tablehead{\colhead{Ion}&{$E^a$}&{$\lambda_{vac}$}&{$g_{low}$}&{log f}&{Ref$^b$}}
\startdata
\cii\   &    0.00  & 1334.53       &  2 & -0.92 & 1 \\  
\cii*   &   63.42  & 1335.70$^c$ &  4 & -0.92 & 1 \\
\civ\   &    0.00  & 1548.19       &  2 & -0.72 & 1 \\
\civ\   &    0.00  & 1550.77       &  2 & -1.02 & 1 \\
\mgii\  &    0.00  & 2796.36       &  2 & -0.20 & 1 \\
\mgii\  &    0.00  & 2803.54       &  2 & -0.51 & 1 \\
\alii\  &    0.00  & 1670.79       &  1 &  0.32 & 1 \\
\aliii\ &    0.00  & 1854.72       &  2 & -0.24 & 1 \\
\aliii\ &    0.00  & 1862.79       &  2 & -0.54 & 1 \\
\siII\  &    0.00  & 1304.37       &  2 & -1.04 & 2 \\
\siII*  &  287.24  & 1309.28       &  4 & -1.09 & 2 \\
\siII\  &    0.00  & 1526.71       &  2 & -0.88 & 2 \\
\siII*  &  287.24  & 1533.43       &  4 & -0.89 & 2 \\
\siII\  &    0.00  & 1808.01       &  2 & -2.60 & 2 \\
\siII*  &  287.24  & 1816.93$^c$ &  4 & -2.67 & 2 \\
\siIV\  &    0.00  & 1393.76       &  2 & -0.27 & 1 \\
\siIV\  &    0.00  & 1402.77       &  2 & -0.58 & 1 \\
\feii\  &    0.00  & 2344.21       & 10 & -0.94 & 3 \\
\feii\  &    0.00  & 2382.77       & 10 & -0.50 & 3 \\
\feii*  &  384.79  & 2396.36       &  8 & -0.54 & 3 \\
\feii\  &    0.00  & 2586.65       & 10 & -1.16 & 3 \\
\feii\  &    0.00  & 2600.17       & 10 & -0.62 & 3 \\
\enddata
\tablenotetext{a} { Energy level in cm$^{-1}$}
\tablenotetext{b} { References: 1--\citet{Kurucz1995},  2--\citet{Tayal2007} and 3--\citet{Morton2003}}
\tablenotetext{c} { Blend of two transitions}
\end{deluxetable}

\begin{deluxetable}{lccccccc}
\label{columndensities}
\tablecaption{Comparison of Column Densities Using the Apparent Optical Depth
and Inhomogeneous Absorber Models}
\tablehead{\colhead{Ion}&{E$^{a}$}&{log $N$(Ap)$^{b}$}&{\siIV\ $a$ log $N$(Inh)$^{c}$}&{\aliii\ $a$ log $N$(Inh)$^{c}$}&{Adopted$^{d}$ log $N$}}
\startdata
\cii\ \& \cii*$^e$ & 0 \& 63 & 14.88  &  14.92$\pm$0.03  &  14.94$\pm$0.03  &  14.93$\pm$0.04  \\  
\civ\          &    0    & 15.66  &  16.20$\pm$0.06  &  16.54$\pm$0.17  &  16.42$\pm$0.28  \\
\mgii\         &    0    & 13.85  &  13.87$\pm$0.08  &  13.89$\pm$0.08  &  13.88$\pm$0.09  \\
\alii\         &    0    & 12.88  &  12.89$\pm$0.09  &  12.91$\pm$0.09  &  12.90$\pm$0.10  \\
\aliii\        &    0    & 14.38  &  14.52$\pm$0.04$^f$  &  14.60$\pm$0.05  &  14.56$\pm$0.08  \\
\siII\         &    0    & 13.88  &  13.89$\pm$0.04  &  13.91$\pm$0.04  &  13.90$\pm$0.05  \\
\siII*         &  287    & 14.07  &  14.08$\pm$0.04  &  14.10$\pm$0.04  &  14.09$\pm$0.05  \\
\siIV\         &    0    & 15.47  &  15.93$\pm$0.04  &  16.24$\pm$0.16$^f$  &  16.14$\pm$0.24  \\ 
\feii\         &    0    & 12.65  &  12.65$\pm$0.14  &  12.67$\pm$0.14  &  12.66$\pm$0.15  \\
\feii*         &  385    & 12.15  &  12.15$\pm$0.29  &  12.16$\pm$0.29  &  12.15$\pm$0.30  \\
\enddata
\tablenotetext{a} { Energy level in cm$^{-1}$.}
\tablenotetext{b} { Apparent column density in log units of cm$^{-2}$, statistical errors are smaller than errors in inhomogeneous absorber models.}
\tablenotetext{c} { Inhomogeneous absorber column density using the indicated power-law $a$ solution in log units of cm$^{-2}$.}
\tablenotetext{d} {The adopted column densities are derived such they span the values and 1$\sigma$ errors of the measurements obtained using the two power-law $a$ models, i.e. Columns 4 and 5.}
\tablenotetext{e} { Assuming N(E=0)/N(E=63 cm$^{-1}$) is the statistical weight ratio of 1:2, see text.}
\tablenotetext{f} { Average template power-law $a$ solution actually used.}
\end{deluxetable}

\begin{deluxetable}{lccccccc}
\label{photoiontab}
\tabletypesize{\scriptsize}
\tablecaption{Photoionization Models for Outflow Component $c$ in SDSS~J0838+2955} 
\tablehead{
\colhead{Parameter} &
\colhead{SED} &
\colhead{MF87}&
\colhead{mod MF87} &  
\colhead{H02} &
\colhead{MF87} & 
\colhead{mod MF87} &  
\colhead{H02} \cr
\colhead{} &
\colhead{Constraint}&
\colhead{Carbon}&
\colhead{Carbon} &  
\colhead{Carbon} &
\colhead{Silicon} & 
\colhead{Silicon} & 
\colhead{Silicon} \cr
\colhead{} &
\colhead{$\log(N)$ (cm$^{-2}$)}&
\colhead{Solar}&
\colhead{Solar} &  
\colhead{Solar} &
\colhead{Solar} & 
\colhead{Solar} & 
\colhead{Solar}
}
\startdata 
$\log(N_H)    $&$           $ &   20.50 &   20.54 &   20.54 &   21.13 &   21.14 &   21.28  \cr
$\log(U_H)    $&$           $ &  --2.12 &  --2.16 &  --2.02 &  --1.68 &  --1.73 &  --1.57  \cr
$\log(T_e)    $&$           $ &    4.08 &    4.13 &    4.13 &    4.08 &    4.13 &    4.11  \cr
$\chi^2_{\nu} $&$           $ &     4.2 &      14 &    67   &    9.6  &   9.2   &   37     \cr   
\cii          &$14.93\pm0.04$ &    0.00 &    0.00 &    0.00 &    0.13 &    0.14 &    0.32  \cr
\civ          &$16.42\pm0.28$ &    0.00 &    0.00 &    0.00 &    0.78 &    0.72 &    0.84  \cr
\mgii         &$13.88\pm0.09$ &    0.23 &  --0.01 &  --0.20 &    0.77 &    0.69 &    0.73  \cr
\alii         &$12.90\pm0.10$ &    0.24 &    0.06 &  --0.10 &    0.39 &    0.38 &    0.46  \cr
\aliii        &$14.56\pm0.08$ &  --0.79 &  --1.03 &  --1.13 &  --0.32 &  --0.43 &  --0.24  \cr
\siII         &$14.33\pm0.05$ &  --0.11 &  --0.24 &  --0.56 &    0.00 &    0.00 &    0.00  \cr
\siIV         &$16.14\pm0.24$ &  --0.47 &  --0.50 &  --0.68 &    0.00 &    0.00 &    0.00  \cr
\feii         &$13.03\pm0.17$ &  --1.14 &  --1.22 &  --1.34 &  --1.14 &  --1.10 &  --1.02  \cr
\enddata
\tablenotetext{Notes.} { The first column is the measured ionic column densities while successive columns are the predicted minus measured ionic column densities based on the solution for that particular model. All quantities are in log representation.  Solar abundances have been used for these six models, and the chosen SED is indicated in the top row for each model.   The first three predicted column densities are the solutions to \cii\ and \civ\ and the last three columns are the solutions to \siII\ and \siIV. The listed temperature is the average $T_e$ across the entire slab.  The $\chi^2_{\nu}$ statistic is also reported although it was never minimized to determine the solution to the pair of ions.  The  $\chi^2_{\nu}$ value is determined only from the \cii, \civ, \siII, and \siIV\ values.} 
\end{deluxetable}

\begin{deluxetable}{lccccccc}
\tabletypesize{\scriptsize}
\tablehead{
\colhead{Parameter} &
\colhead{SED} &
\colhead{MF87} & 
\colhead{mod MF87} &  
\colhead{H02} &
\colhead{MF87} & 
\colhead{mod MF87} &  
\colhead{H02} \cr
\colhead{} &
\colhead{Constraint}&
\colhead{C \& Si} &
\colhead{C \& Si} &
\colhead{C \& Si} &
\colhead{C \& Si} &
\colhead{C \& Si} &
\colhead{C \& Si} \cr
\colhead{} &
\colhead{$\log(N)$ (cm$^{-2}$)}& 
\colhead{C: --0.09} &
\colhead{C: --0.15} &
\colhead{C: --0.32} &
\colhead{Si: +0.14} &
\colhead{Si: +0.26} &
\colhead{Si: +0.60}
}
\startdata 
$\log(N_H)    $&$           $ &   20.73 &   20.84 &   21.03 &   20.63 &   20.60 &   20.60 \cr
$\log(U_H)    $&$           $ &  --1.97 &  --1.97 &  --1.77 &  --2.01 &  --2.11 &  --1.97 \cr
$\log(T_e)    $&$           $ &    4.08 &    4.13 &    4.10 &    4.09 &    4.14 &    4.13 \cr
$\chi^2_{\nu} $&$           $ &    1.8  &    1.9  &    1.2  &    1.5  &   0.93  &    0.23 \cr   
\cii          &$14.93\pm0.04$ &    0.00 &  --0.01 &    0.00 &  --0.01 &  --0.01 &  --0.01 \cr
\civ          &$16.42\pm0.28$ &    0.21 &    0.22 &    0.22 &    0.20 &    0.11 &    0.08 \cr
\mgii         &$13.88\pm0.09$ &    0.49 &    0.48 &    0.58 &    0.31 &    0.04 &  --0.18 \cr
\alii         &$12.90\pm0.10$ &    0.35 &    0.35 &    0.46 &    0.22 &    0.06 &  --0.11 \cr
\aliii        &$14.56\pm0.08$ &  --0.58 &  --0.65 &  --0.40 &  --0.72 &  --1.00 &  --1.10 \cr
\siII         &$14.33\pm0.05$ &    0.00 &    0.01 &    0.01 &    0.01 &    0.01 &    0.01 \cr
\siIV         &$16.14\pm0.24$ &  --0.27 &  --0.26 &  --0.18 &  --0.23 &  --0.20 &  --0.05 \cr
\feii         &$13.03\pm0.17$ &  --1.07 &  --1.01 &  --0.93 &  --1.20 &  --1.25 &  --1.37 \cr
\enddata
\tablenotetext{Notes.} { Similar to Table 3 except the solutions are determined by minimizing the $\chi^2_{\nu}$ statistic with the option to vary the chemical abundance of an element.  The first set of three models allow the carbon abundance to decrease while the last three models allow the silicon abundance to increase.  The amount that the chemical abundance was varied from solar is indicated in the third row.} 
\end{deluxetable}

\begin{figure}
\centering
\includegraphics[width=4.3in]{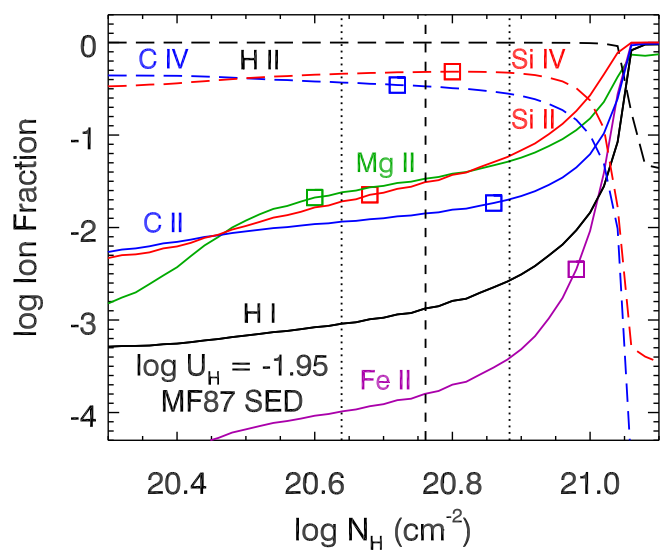}
\caption{Representative slab model of absorber component $c$ assuming the MF87 SED, log $U_H$ = -1.95 and chemical abundances where carbon is decreased by 0.3 dex relative to solar abundances while silicon is increased by 0.2 dex.  The squares indicate the total $N_H$ necessary so that the predicted integrated ionic column densities match the measured ones.  The vertical lines represent the adopted value (dashed) and error (dotted) of $N_H$ for the given $U_H$.}
\label{cloud.fig}
\end{figure}

\end{document}